\documentclass[twocolumn,prl,showpacs,floatfix]{revtex4}
\usepackage{graphicx}
\usepackage{epsfig}
\usepackage{rotating}
\usepackage{amsmath}
\usepackage{amsfonts}
\usepackage{amssymb}
\usepackage{enumerate}
\usepackage{longtable}
\usepackage{dcolumn}
\usepackage{bm}

\begin{document}

\title{High temperature thermodynamics of the honeycomb-lattice Kitaev-Heisenberg model: A high temperature series expansion study}

\author{R. R. P. Singh}
\affiliation{University of California Davis, CA 95616, USA}

\author{J. Oitmaa }
\affiliation{School of Physics, The University of New South Wales,
Sydney 2052, Australia}

\date{\rm\today}

\begin{abstract}
We develop high temperature series expansions for the thermodynamic properties of the honeycomb-lattice
Kitaev-Heisenberg model. 
Numerical results for uniform susceptibility, heat capacity and entropy as a function of temperature
for different values of the Kitaev coupling $K$ and Heisenberg exachange coupling $J$ (with $|J|\le |K|$)  are presented.
These expansions show good convergence down to a temperature of a fraction of $K$ and in some cases down to $T=K/10$.
In the Kitaev exchange dominated regime, the inverse susceptibility has a nearly linear temperature dependence 
over a wide temperature range.
However, we show that already at temperatures $10$-times the Curie-Weiss temperature, 
the effective Curie-Weiss constant estimated from the data can be off by a factor of 2.
We find that the magnitude of the heat capacity maximum at the short-range
order peak, is substantially smaller for small $J/K$ than for $J$ of order or larger than $K$.
We suggest that this itself represents a simple marker for the relative importance of the Kitaev terms
in these systems. Somewhat surprisingly, both heat capacity and susceptibility data on Na$_2$IrO$_3$ are consistent
with a dominant {\it antiferromagnetic} Kitaev exchange constant of about $300-400$ $K$.

\end{abstract}


\maketitle

\section{Introduction}
Kitaev's discovery of a class of exactly soluble honeycomb-lattice, anisotropic, spin-half models with gapped
and gapless spin-liquid phases and Majorana fermion excitations \cite{kitaev}, represents a major advance in the field of 
quantum magnetism.
Furthermore, Jackeli and Khaliullin's demonstration that such special Kitaev-couplings can indeed be realized
in real materials \cite{jk} has led to intense theoretical and experimental activity in the field. Several
classes of materials dubbed `Kitaev-materials'\cite{trebst,valenti,rau} have been synthesized and 
these now represent some of
the most promising candidates for the much sought after quantum spin-liquid phase of matter \cite{balents-review}.
Several
neutron scattering and other experimental studies have been interpreted as evidence for proximate spin-liquid behavior,
even though the true ground-state may often be long-range ordered \cite{neutron1,neutron2,knolle1,knolle2,analytis}.

Heisenberg exchange couplings are the most generic terms in modeling quantum magnetism. Even 
in Kitaev-materials, there is always varying degree of Heisenberg exchange couplings present, which can affect
key properties including driving the system away from the spin-liquid phases and into various magnetically
ordered phases \cite{chaloupka1,chaloupka2,reuther,jiang,kimchi,schaffer,song,price,motome,nasu,duality,halasz}. 
We could also take a phenomenological point of view that in the Kitaev materials, deviations from the Kitaev model can be subsumed into effective Heisenberg couplings.
These Heisenberg couplings destroy the exact solubility of the model and necessitate
numerical studies. Previously these models have been studied numerically by Monte Carlo simulations, Exact Diagonalization and
other techniques based on finite-size clusters \cite{chaloupka1,chaloupka2,reuther,jiang,kimchi,schaffer,song,price,motome,nasu,duality,halasz}.

Here we present a study of the nearest-neighbor Kitaev-Heisenberg model using the high temperature series expansion (HTSE) method \cite{book,advphy}.
These series expansions are formally defined in the thermodynamic limit and give accurate properties of the infinite system 
when the expansions are convergent typically at temperatures above the exchange energy scales. At lower temperatures, one can use Pade extrapolation methods to obtain the
desired thermodynamic properties. We typically find good convergence up to and a little below the temperature of
the peak in the heat capacity associated with short-range magnetic order in the system.

The uniform susceptibility is often used to obtain the first estimates of the exchange constants for magnetic materials.
In the Kitaev coupling dominated regime, the inverse susceptibility appears to be nearly linear over a wide
temperature range. Plots of inverse susceptibility versus temperature\cite{ys1,ys2,ys3} can be extrapolated
to obtain the Curie-Weiss constant. However, one needs to be careful in relating these to the microscopic exchange
constants for these systems as already at temperatures $10$ times the Curie-Weiss
temperature the effective Curie-Weiss constant obtained from such an extrapolation can be off by a factor of 2.
\begin{figure}
\begin{center}
 \includegraphics[angle=270,width=0.8\columnwidth]{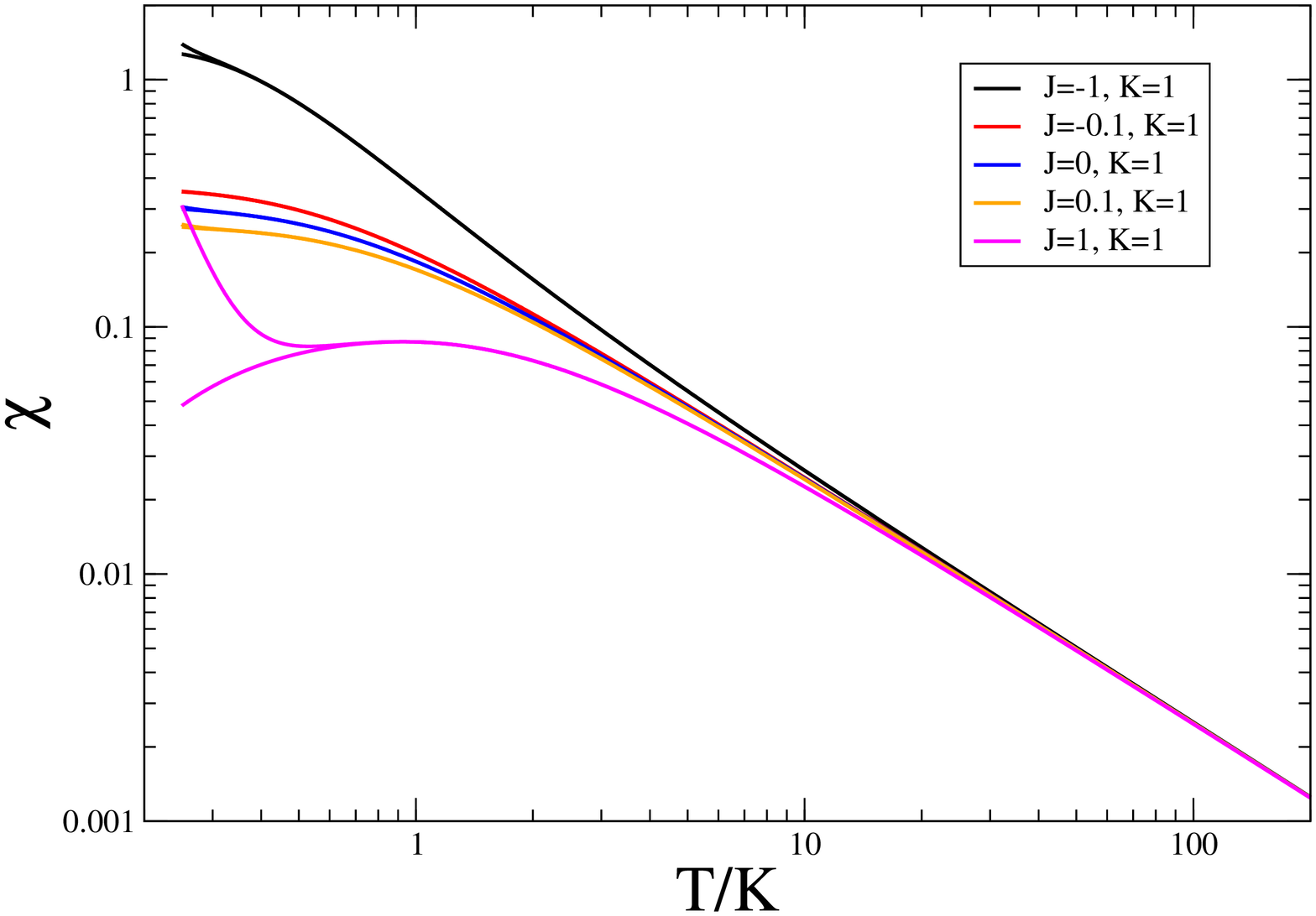}

 \includegraphics[angle=270,width=0.8\columnwidth]{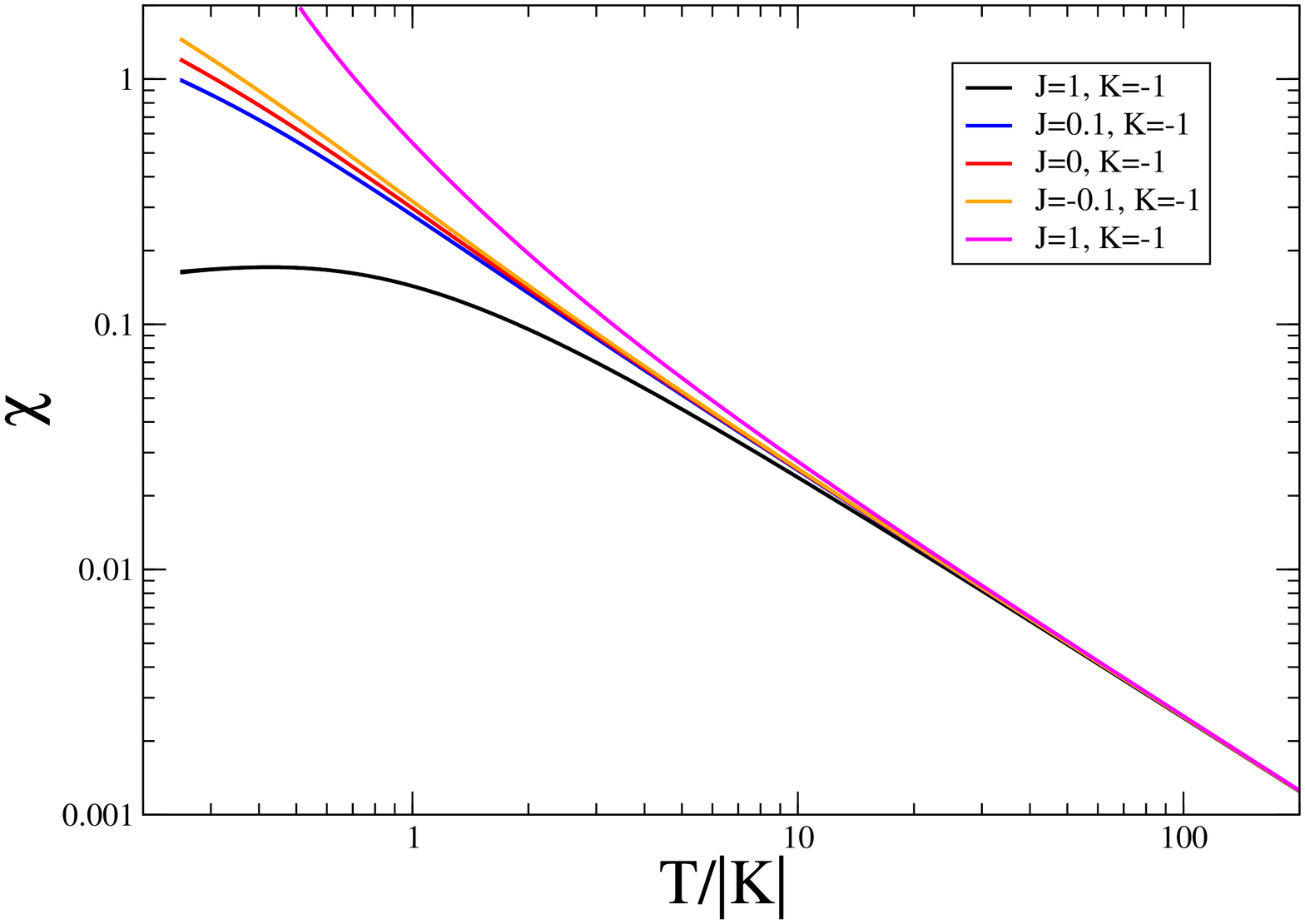}
\caption{\label{fig1} 
A plot of the uniform susceptibility of the model as a function of temperature $T$ for different values of 
exchange constants $K$ and $J$. 
}
\end{center}
\end{figure}

The behavior of entropy and heat capacity in the Kitaev coupling dominated regime is qualitatively
different from that in the Heisenberg coupling dominated regime and can serve as a simple marker
for the importance of Kitaev couplings in a real material. In this regime, the entropy is realeased
only partially at the onset of short-range order marking the beginning of a proximate Kitaev spin-liquid
intermediate temperature regime. In their Monte Carlo studies, Nasu et al \cite{nasu} found
a plateau in the entropy over a range of temperatures at a value of ${1\over 2} R \ln{2}$. We are studying
a model in which the Kitaev couplings are same along the three bond-directions of the Honeycomb lattice,
where there may not be such a strict plateau. Our studies show
only hints of a possible plateau formation not different from what is seen in some experiments \cite{ys3},
with the actual plateau region being at still lower temperatures than reached in our studies and preempted
by long-range order in experiments.

Quite remarkably, the magnitude of the short-ranged peak in the heat capacity is itself a marker of whether
the Kitaev terms are important or whether the system is dominated by Heisenberg exchange terms.
In the Kitaev regime this peak is of significantly smaller magnitude.
We suggest that this may itself provide a good experimental means to quickly characterize which materials are
likely to be near the quantum spin-liquid phase.

\section{Models}
We study the honeycomb-lattice Kitaev-Heisenberg model with Hamiltonian
\begin{equation}
{\cal H}=J \sum_{<i,j>} \vec S_i \cdot \vec S_j
+K \sum_{<i,j>_\alpha} S_i^\alpha S_j^\alpha,
\end{equation}
where the sums run over all nearest neighbor pairs of the Honeycomb lattice. The second
sum represents the Kitaev couplings. The bonds of the honeycomb lattice can be divided into
3 different types labelled by $\alpha=$ $x$, $y$, or $z$. The Kitaev coupling along the bond type $\alpha$
involves bilinear spin-couplings involving only the spin operator $S_i^\alpha$. We note that, in the literature,
the opposite sign of the couplings as well as a factor of 2 in the definition of the Kitaev terms has
sometimes been used. And, in fact, theory points towards a ferromagnetic Kitaev exchange, which in our notation
corresponds to negative $K$. Also, in our study the strength of the Kitaev couplings are the same in all directions.

\begin{figure}
\begin{center}
 \includegraphics[angle=270,width=0.8\columnwidth]{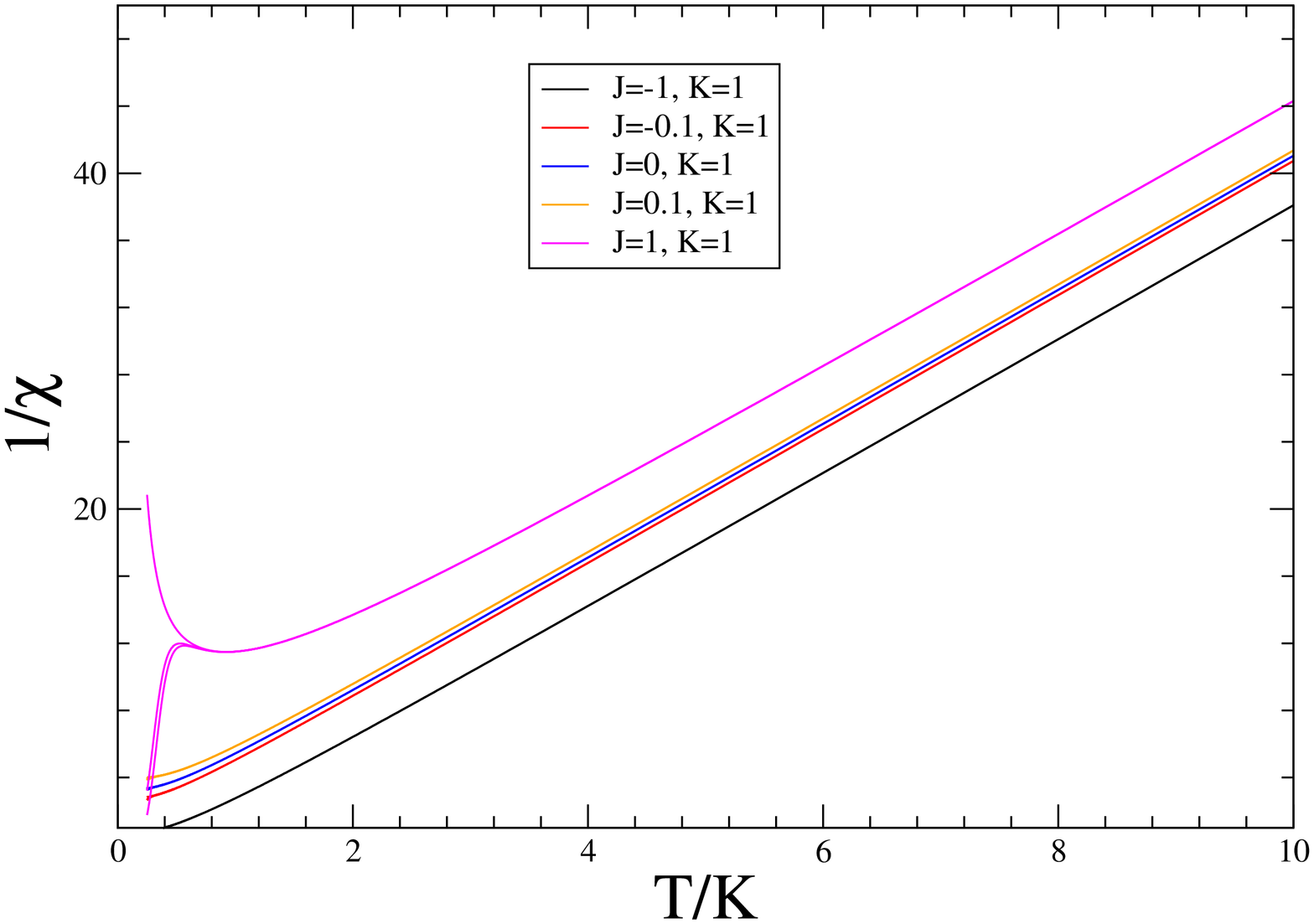}

 \includegraphics[angle=270,width=0.8\columnwidth]{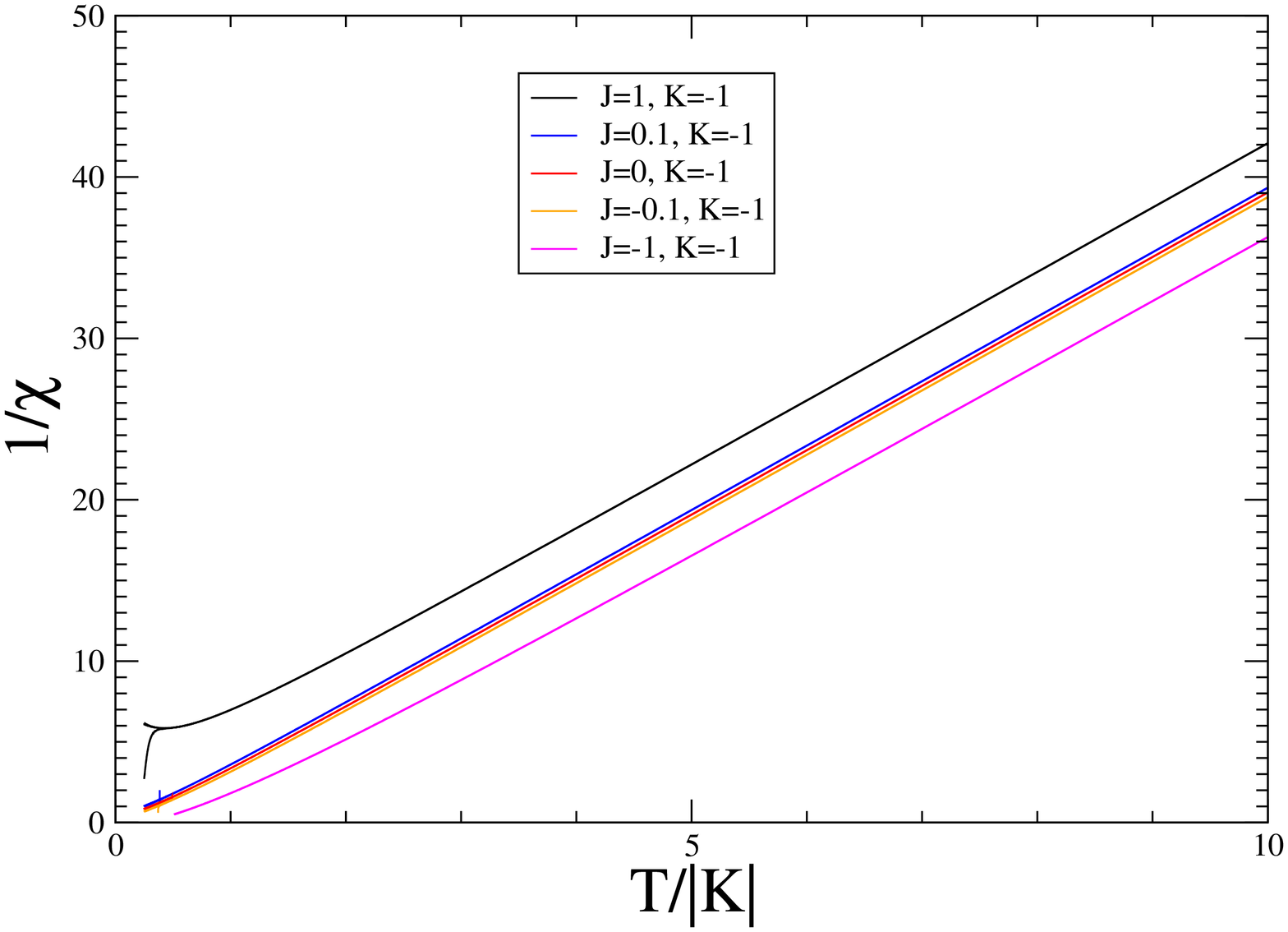}
\caption{\label{fig2} 
A plot of the inverse susceptibility of the model as a function of temperature $T$ for different values of 
exchange constants $K$ and $J$. The linear temperature dependence of the inverse susceptibility seemingly extends
well below $T=K$.
}
\end{center}
\end{figure}

\begin{figure}
\begin{center}
 \includegraphics[angle=270,width=\columnwidth]{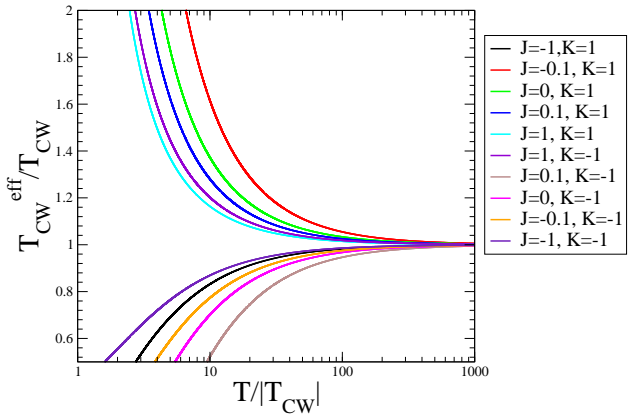}
\caption{\label{fig3} 
The effective Curie-Weiss $T^{eff}_{cw}$ constant obtained by a straight-line extrapolation of the inverse susceptibility
locally from some temperature $T$ as a function of temperature $T$.
Note that in the Kitaev regime the effective Curie-Weiss constant can differ from the true Curie-Weiss $T_{cw}$
constant by a factor of two already at a temperature $10$ times the Curie-Weiss temperature.
}
\end{center}
\end{figure}

We develop high temperature series expansion for the logarithm of the partition function using the linked-cluster
method, from which the internal energy, heat capacity and entropy follows. These calculations are done
to order $\beta^{14}$. Series coefficients for selected values of $J$ and $K$ are given in supplementary materials.
In addition, we also apply an external-field along the x-axis, and calculate the uniform susceptibility as
the second derivative of the free-energy with respect to this external field. These expansions are carried
out for $T\chi$ to order $\beta^{12}$. The series coefficients for the susceptibilities are also given
in the supplementary materials. We will present results for ferromagnetic and antiferromagnetic Kitaev coupling $K$
as well as for $J/K$ ratios of $0$, $\pm 0.1$ and $\pm 1$ to cover a wide range of behaviors. Larger $J$ values are closer
to the $J/K=1$ behavior so that the system is well in the Heisenberg dominated regime already when the two couplings are equal.
We call the parameter range $|J/K|\le 0.1$ as the Kitaev regime. At high temperatures the thermodynamic 
properties are not so sensitive to much smaller variations in parameters.

\section{Results}
We analyze the series by using Pade approximants. As long as different approximants agree with each other,
those usually imply convergence in the thermodynamic limit. At least two approximants are plotted
in each case. Some times Pade approximants
have nearby pole-zero pairs around some temperature. 
That spoils the convergence around that temperature and shows up as a sharp
glitch in our plots. Such a glitch need not affect the convergence of the approximants away from the neighborhood
of that temperature.
However, at low temperatures different approximants clearly start diverging from each other. Then, the extrapolations
can no longer be relied upon.

We begin in Fig.~1 with a plot of the uniform susceptibility as a function of temperature on a log-log
scale. We see that in the Kitaev regime  the plots remain close to each other. Note
that for $J=0$, even though the entropy and heat capacity do not depend on the sign of $K$, the susceptibility
very much does.

The asymptotic Curie-Weiss constant for the model can be obtained from the very first order of the high temperature
expansion as:
\begin{equation}
T_{cw}=0.75 J + 0.25 K.
\end{equation}
In Fig.~2, we show the Curie-Weiss plot of inverse susceptibility versus temperature. In the Kitaev regime,
the plot looks linear over a wide temperature range. Following Kouvel and Fisher \cite{kouvel,zheng} one can
define an effective Curie-Weiss temperature as
\begin{equation}
T^{eff}_{cw}=-T-{\chi\over d\chi/dT}
\end{equation}
If we draw a straight-line to the inverse susceptibility plot at some temperature $T$, 
$T^{eff}_{cw}$ would be the intercept.
In other words, if one obtains Curie-Weiss constant from experimental data up to some temperature only an
effective Curie-Weiss constant will result. In Fig~3, we show the effective Curie-Weiss constant as
a function of temperature. The temperature
scale itself is normalized by the magnitude of the Curie-Weiss constant for the given exchange values. Note that
when the Curie-Weiss constant is antiferromagnetic, the effective Curie-Weiss constant increases in magnitude as temperature
goes down whereas when the Curie-Weiss constant is ferromagnetic it decreases in magnitude as the temperature is lowered.
The Kitaev regime shows very substantial deviations from the high temperature behavior already at a temperature
$10$ times $T_{cw}$. This needs to be taken into account in obtaining exchange constants from such measurements.

\begin{figure}
\begin{center}
 \includegraphics[angle=270,width=0.8\columnwidth]{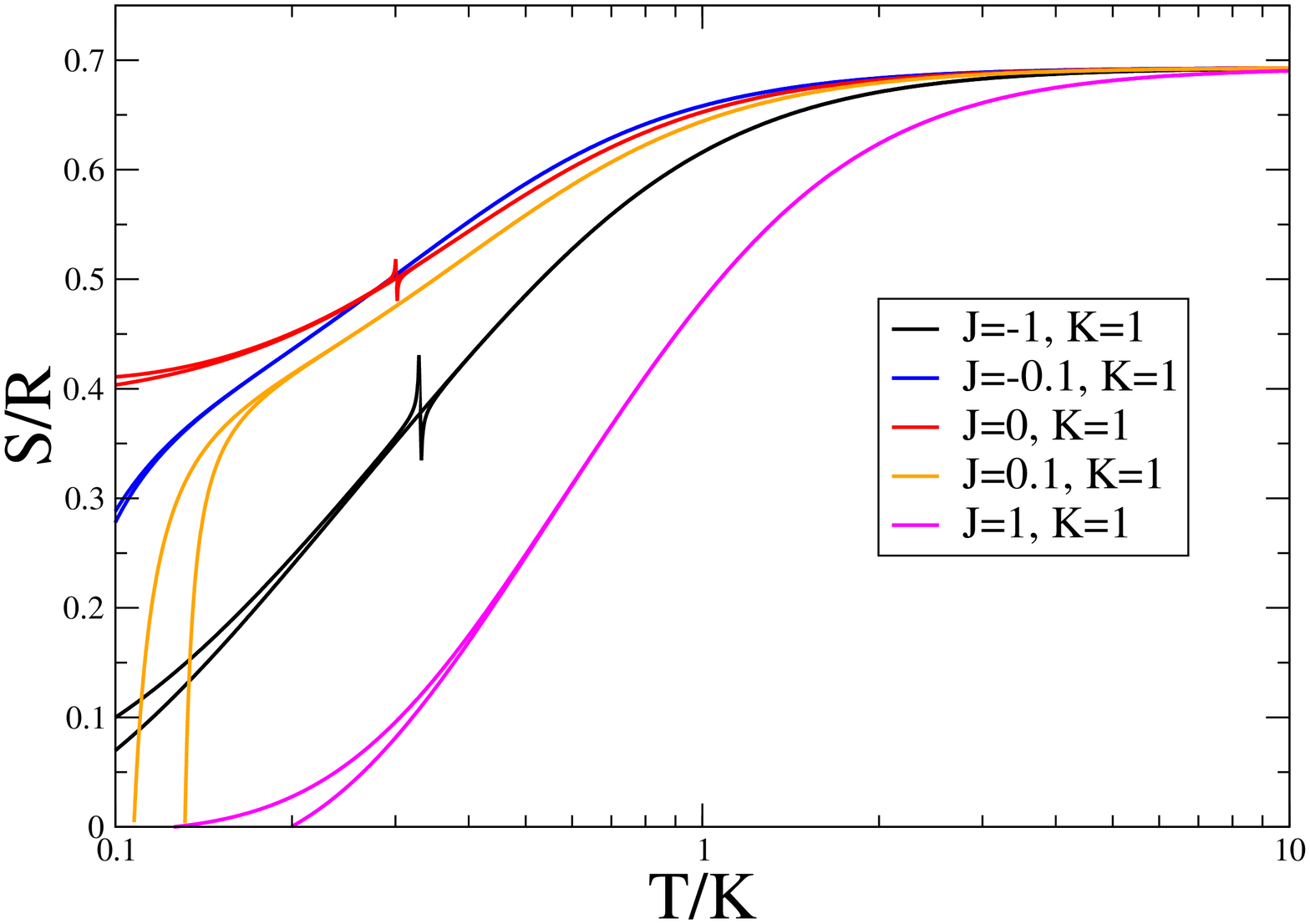}

 \includegraphics[angle=270,width=0.8\columnwidth]{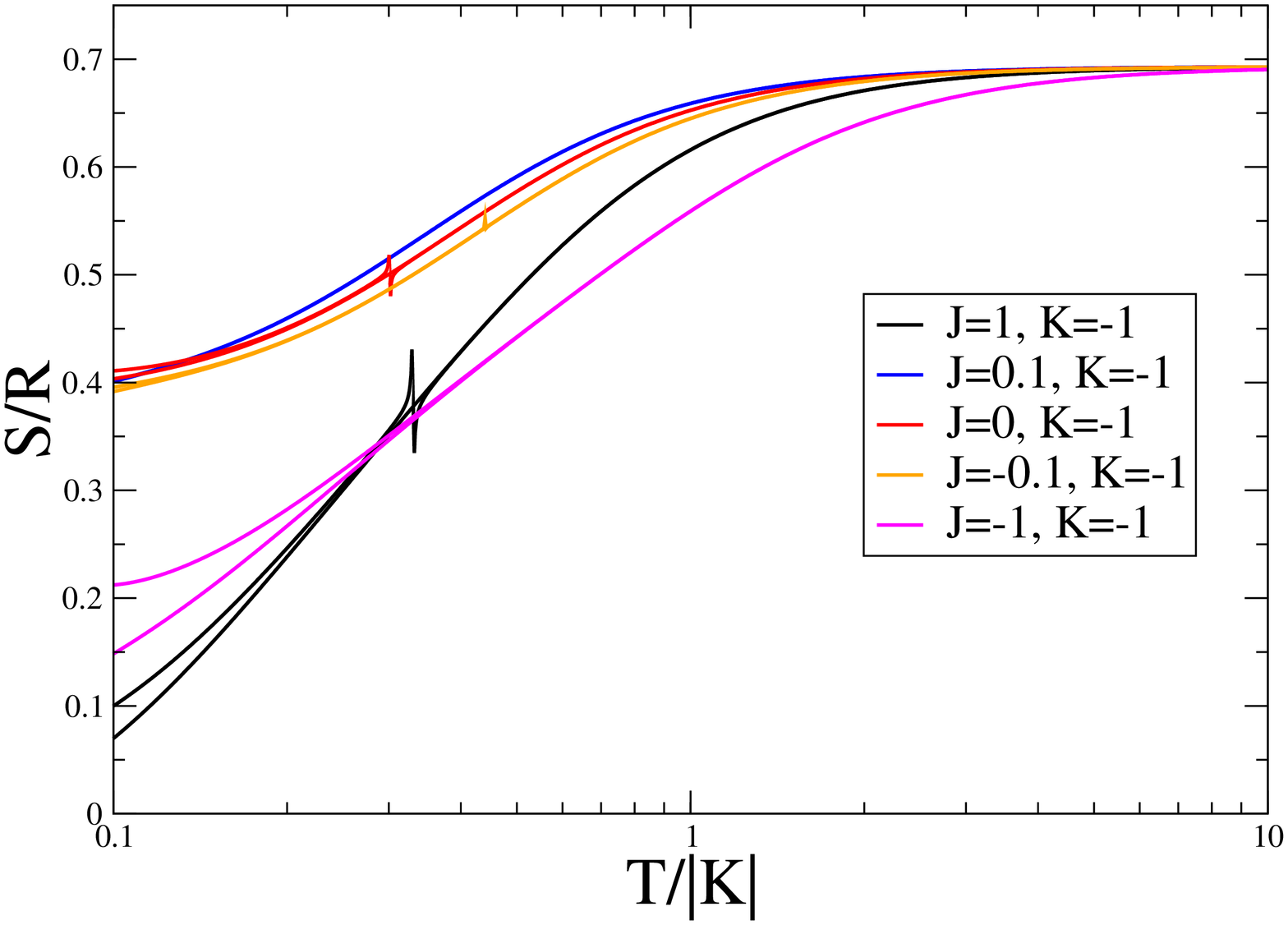}
\caption{\label{fig4} 
A plot of the entropy per mole of the model as a function of temperature $T$ for different values of the
Heisenberg coupling $J$ and the Kitaev coupling $K$.
}
\end{center}
\end{figure}

\begin{figure}
\begin{center}
 \includegraphics[angle=270,width=0.8\columnwidth]{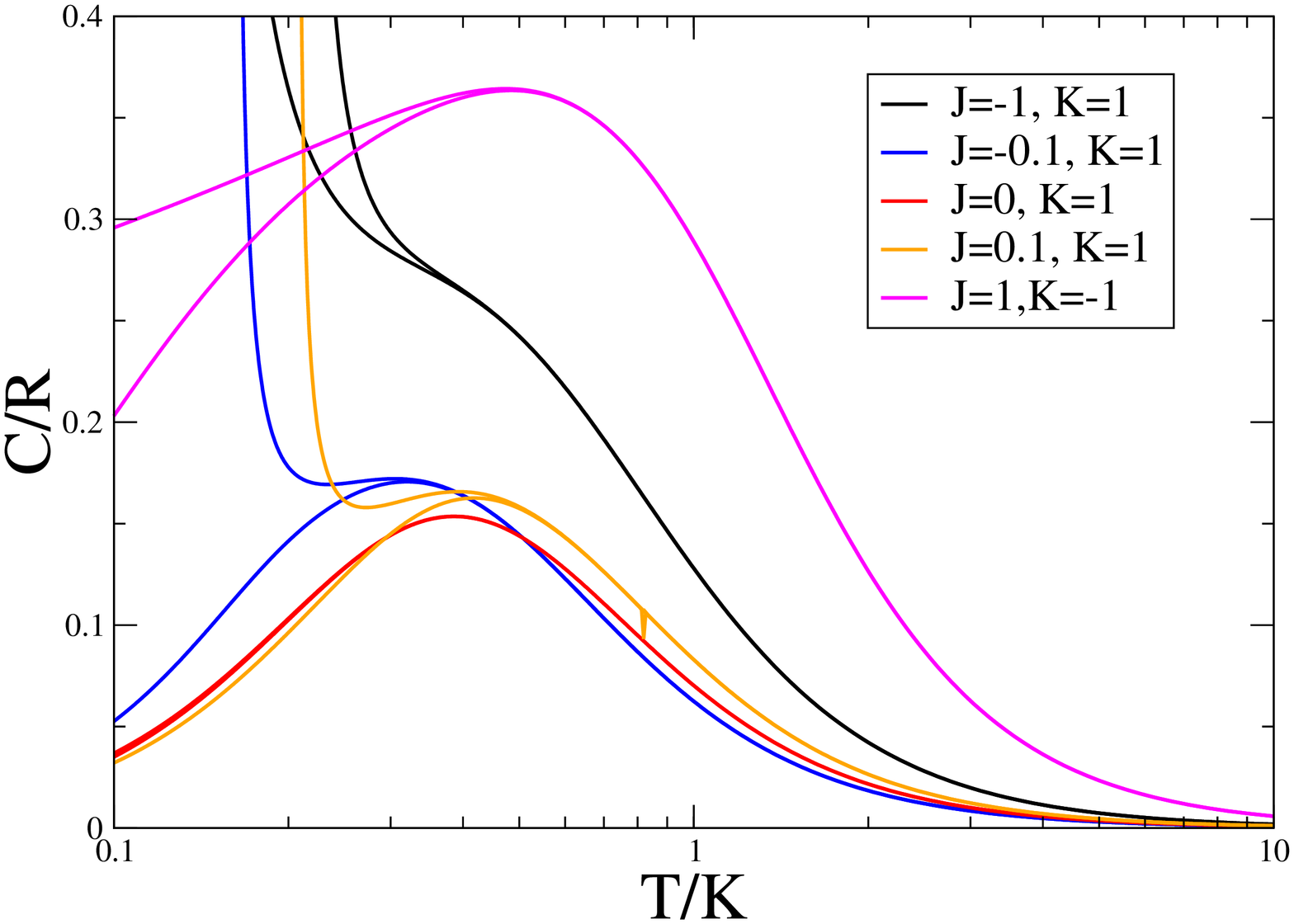}

 \includegraphics[angle=270,width=0.8\columnwidth]{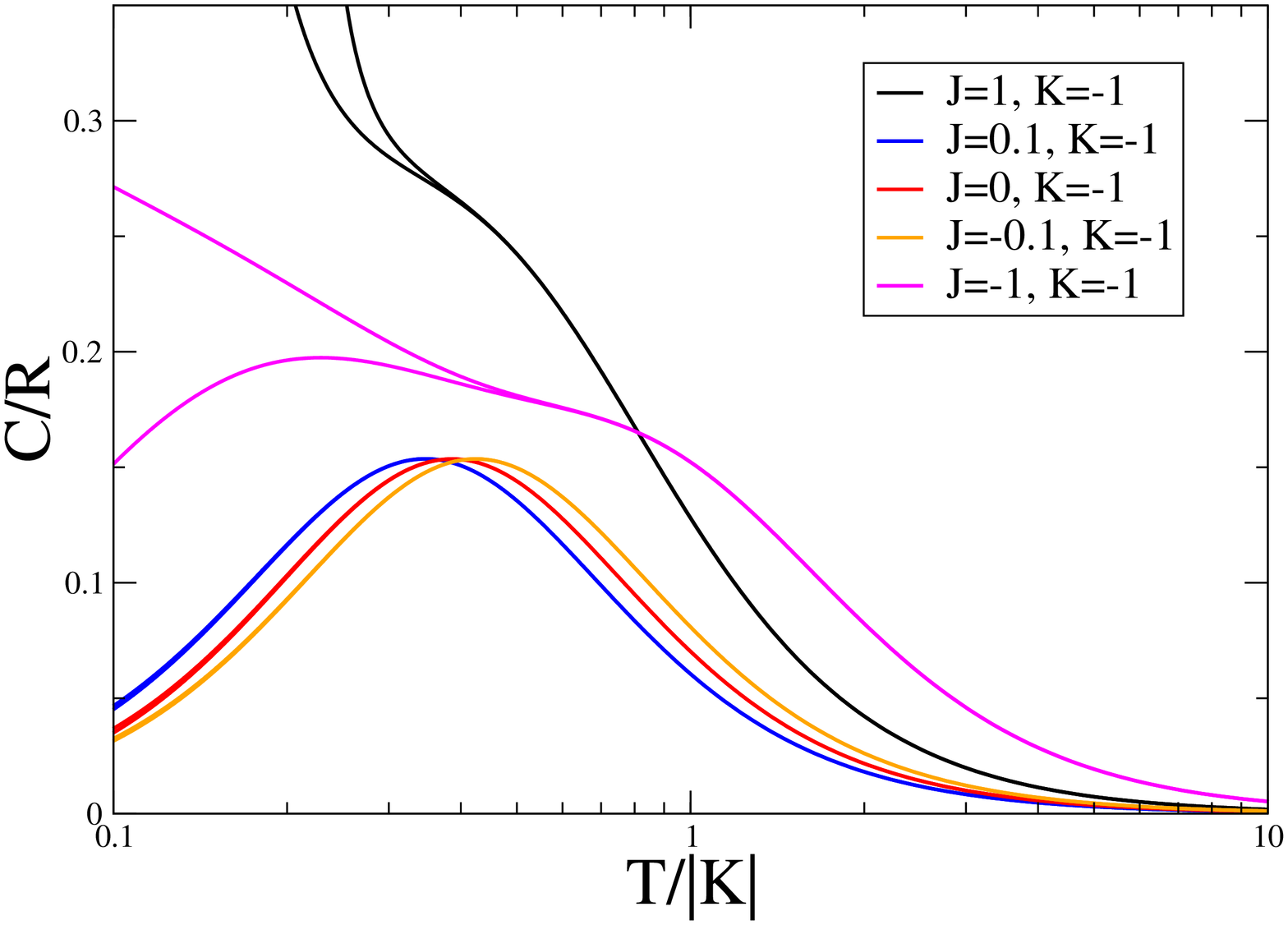}
\caption{\label{fig5} 
A plot of the heat capacity per mole of the model in units of the gas constant $R$
as a function of temperature $T$ for different values of the
Heisenberg coupling $J$ and the Kitaev coupling $K$.
}
\end{center}
\end{figure}

Plots of the molar entropy as a function of temperature is shown in Fig.~4 and the heat capacity in Fig.~5,
with the temperature on a logarithmic scale.
The main message in these plots is very simple. The Kitaev regime corresponding to
small $J/K$ has very different temperature dependence of heat capacity and entropy than
Heisenberg systems where most of the entropy is released in the development of short-range order. 
That is not the case in the Kitaev regime.
Nasu et al have emphasized a plateau in the entropy at a value of $ln{2}/2$
in the gapped spin-liquid models and a plateau-like feature in the gapless spin-liquid models \cite{nasu}.
This is consistent with the idea of
a two step release of entropy and the development of an intermedaite proximate spin-liquid regime.

We are studying here the gapless spin-liquid model only as the Kitaev couplings are taken to be
equal along all three bond-directions of the honeycomb lattice. We do not see a plateau in the entropy plots. 
But, for the pure Kitaev model and a range
of Heisenberg couplings the development of a plateau-like feature at still lower temperature is plausible.
Note that the entropy at the lowest temperatures we show is still above $1/2$ $\ln{2}$.
This feature seems more robust when the Kitaev exchange is ferromagnetic than when it is antiferromagnetic.
Hence, we would conclude that if a plateau actually develops that would be at temperatures below $T/|K|=0.1$.

Pade extrapolations for the heat capacity are shown in Fig.~5.
A striking thing about the plots is
that the magnitude of the heat capacity peak at high-temperatures where the short-range
order develops is significantly smaller in the Kitaev regime than that when the system is dominated by the Heisenberg
coupling and is likely to be deep in the long-range ordered phase at $T=0$. Although 
our results are not always well converged below the short-range order peak, we can clearly say that the
magnitude of this peak remains roughly unchanged in the Kitaev regime. 

In the absence of a finite temperature phase transition, it should be possible, in principle, to extrapolate
thermodynamic properties all the way down to $T=0$. However, 
that usually does not work well \cite{book,elstner}.
Bernu and Misguich have suggested a very different extrapolation procedure for the heat capacity \cite{bernu} that takes
advantage of known zero-temperature properties. While this may be helpful for the pure Kitaev model, low temperature
properties of the infinite system are not known well enough for the Kitaev-Heisenberg model. And, the extrapolations
can be highly sensitive to the exact  $T=0$ properties as well as to the full details of the model parameters. 
We leave such extrapolations for a later study. 

We turn now to experimental systems.
Various researchers have discussed the possibility of exchange constants other than Kitaev and Heisenberg
terms as well as various anisotropies and further neighbor
terms in the Hamiltonian. The high temperature thermodynamics are not very sensitive to the full details of
the model. Looking at the data of Mehlawat et al \cite{ys3}, for Na$_2$IrO$_3$ the peak in the heat capacity
occurs at a temperature of about $110$ $K$. In the Kitaev regime, the peak occurs at a temperature of $0.3$ to
$0.4$ in units of the Kitaev coupling. That translates into a Kitaev coupling of about $275-370$ $K$. 
So far, the Kitaev couplings can have either sign. The Curie-Weiss
fit gives an antiferromagnetic Curie-Weiss temperature of about $125$ $K$. This value of the Curie-Weiss constant
would imply an antiferromagnetic Kitaev exchange of approximately $500$ $K$. However, once we take into account that the measured
Curie-Weiss constant is roughly 1.2-1.8 times too large, once again the Kitaev exchange constant is in
the range of $280-400$ $K$. Thus, the high temperature thermodynamics of this material is in good agreement
with the Kitaev regime, with a single {\it antiferromagnetic} Kitaev coupling. 
The data is similar but more erratic for Li$_2$IrO$_3$, where both quantities are
about 20 percent smaller. These results are surprising as theory points to a ferromagnetic Kitaev exchange.
And, in that case, explaining the susceptibility data requires invoking strong second and third neighbor interactions.



\section{Conclusions}
In conclusion, in this paper we have studied the honeycomb-lattice Kitaev-Heisenberg model using the high temperature series expansion method. 
We have presented results for
thermodynamic properties including the entropy, the heat capacity and the uniform susceptibility
for a range of Kitaev and Heisenberg exchange couplings.
We showed that the usual Curie-Weiss plot of inverse susceptibility looks linear in temperature over a wide
temperature range in the Kitaev regime. But it shows significant deviations from the asymptotic
high temperature behavior, which means one needs to be careful in obtaining exchange constants from such
a Curie-Weiss plot. 
Comparison of our results with experiments on Na$_2$IrO$_3$ gives a dominant 
{\it antiferromagnetic} Kitaev exchange constant of $280-400$ $K$.

The heat capacity and entropy show very distinct behavior in the Kitaev-exchange dominated regime.
The heat capacity peak has a much smaller value in this regime and is effectively constant as $J/K$ is varied.
This itself may be used as a marker in experimental studies 
for an early determination of the importance of Kitaev couplings in the materials.
Furthermore, the entropy is released more gradually as the temperature is
lowered creating a very distinct and characteristic experimental signature. 
While we do not see clear evidence of an entropy plateau, our results are consistent with 
the development of such a plateau-like feature at still lower temperature. Clearly, the
entropy release in these systems is like a two step process, where the higher temperature heat capacity peak
only gives rise to an intermediate proximate spin-liquid
phase at intermediate temperatures.
Somewhat surprisingly, a single {\it antiferromagnetic} Kitaev exchange in the range of $275-400$ $K$ 
gives a reasonable account of both heat capacity
and susceptibility data in sodium and lithium iridates.

\begin{acknowledgements}
This work is supported in part by the US National Science Foundation grant number  DMR-1306048
and by computing resources provided by the Australian (APAC) National Facility
\end{acknowledgements}


\end{document}